\documentclass[aps,prb,amsmath,amssymb,twocolumn,superscriptaddress,showpacs]{revtex4-1}
\usepackage{graphicx}
\usepackage{dcolumn}
\usepackage{bm}
\usepackage{hyperref}

\def\be{\begin{eqnarray}}   
\def\ee{\end{eqnarray}}

\begin{document}

\title{Photovoltaic Effect from the Viewpoint of Time-reversal Symmetry}

\author{Shunsuke~A.~Sato}
\email{ssato@ccs.tsukuba.ac.jp}
\affiliation 
{Center for Computational Sciences, University of Tsukuba, Tsukuba 305-8577, Japan}
\affiliation 
{Max Planck Institute for the Structure and Dynamics of Matter, Luruper Chaussee 149, 22761 Hamburg, Germany}

\date{\today}

\begin{abstract}
We theoretically investigate field-induced charge-transport processes
from the viewpoint of time-reversal symmetry.
We analytically demonstrate that breaking of the time-reversal
symmetry is a necessary condition to induce charge-transport and direct-current
by external fields.
This finding provides microscopic insights into photovoltaic effects
and optical-control of currents.
\end{abstract}


\maketitle

The photovoltaic effect, the conversion of energy from light to electric current,
is an important effect from both fundamental and technological points
of view. Recently, the shift-current, which is one of the mechanisms of the
photovoltaic effect, has been attracting interest as an efficient
energy conversion mechanism and has been intensively investigated theoretically
and experimentally \cite{PhysRevB.61.5337,doi:10.1063/1.1436530,PhysRevLett.109.116601,Zheng2015,PhysRevLett.109.236601,Sotome1929}.
In ultrafast sciences and strong-field physics,
control of electric current by strong light has been investigated with the aim of realizing petahertz
optoelectronics \cite{Schiffrin2012,PhysRevLett.113.087401,Wachter_2015,PhysRevLett.116.197401,PhysRevLett.116.057401,Higuchi2017,mciver2018light,PhysRevB.99.214302}.
In this short note, we theoretically investigate field-induced charge-transport
phenomena from the viewpoint of time-reversal symmetry
in order to provide microscopic insights into the photovoltaic effect
and the optical-control of current.

Here, we consider an $N$-particle system described by the following Schr\"odinger equation,
\be
i\hbar \frac{\partial}{\partial t} \Psi(R,t) = \hat H(t) \Psi(R,t),
\label{eq:tdse}
\ee
where coordinates of $N$-particles are collectively denoted by $R:=\{r_1,\cdots,r_N\}$,
and the Hamiltonian is denoted by $\hat H(t)$.
The charge-transport dynamics is investigated by evaluating the current $J(t)$
whose operator is defined as
\be
\hat J(t) = \frac{q}{i\hbar} \left [ R, \hat H(t) \right ],
\ee
where $q$ is the charge of particles.

Now we prove that the charge-transport is forbidden under the following
three conditions. For the sake of simplicity, we consider time propagation from $t=-T/2$
to $t=T/2$.

\textit{The first condition:} The initial state is time-reversal symmetric,
satisfying
\be
\Psi^*\left (R, -\frac{T}{2}\right ) = e^{i \phi_1}\Psi\left (R, -\frac{T}{2}\right ),
\label{eq:tr-sym-wf}
\ee
where $\phi_1$ is a constant phase.

\textit{The second condition:} The system returns to the initial state after
the time-evolution, satisfying
\be
\Psi\left (R, \frac{T}{2}\right ) = e^{-i\phi_2} \Psi\left (R, -\frac{T}{2}\right ),
\label{eq:adiabatic-wf}
\ee
where $\phi_2$ is a constant phase.

\textit{The third condition:} The Hamiltonian $\hat H(t)$ satisfies the following 
time-reversal symmetry condition:
\be
\left [ H(-t) \Phi(R) \right ]^* = \hat H(t) \Phi^*(R),
\label{eq:tr-sym-pot}
\ee
where $\Phi(R)$ is any complex function.

The first condition, Eq.~(\ref{eq:tr-sym-wf}), guarantees that the initial current,
$J(-T/2)=\int dR \Psi^*(R,-T/2)\hat J(-T/2)\Psi(R,-T/2)$, is zero.
The second condition, Eq.~(\ref{eq:adiabatic-wf}), guarantees that the external field
does not leave any excitations to the system after the perturbation.
The third condition, Eq.~(\ref{eq:tr-sym-pot}), guarantees the time-reversal symmetry
of external fields. For example, if the Hamiltonian has the following form,
\be
\hat H(t) = \frac{1}{2m} \left [
-i\hbar \frac{\partial}{\partial R} - \frac{q}{c}A(t) \right ]^2
+V(R,t),
\ee
with a vector potential $A(t)$ and a scalar potential $V(R,t)$,
the third condition, Eq.~(\ref{eq:tr-sym-pot}), leads to the following requirements,
\be
A(t) &=& -A(-t), \label{eq:sym-vec-pot} \\
V(R,t) &=& V(R,-t). \label{eq:sym-scr-pot}
\ee
Note that linearly-polarized light can satisfy Eq.~(\ref{eq:sym-vec-pot})
and Eq.~(\ref{eq:sym-scr-pot}), while 
circularly- or elliptically-polarized light cannot.

To prove that charge transport is forbidden,
we analyze the relation between the forward and backward time-propagations.
Taking the complex conjugate of Eq.~(\ref{eq:tdse}) and replacing $t$ by $-t$, 
the Schr\"odinger equation can be rewritten as
\be
i\hbar \frac{\partial}{\partial t} \Psi^*(R,-t)= \left [\hat H(-t)
\Psi(R,-t) \right]^*
=
\hat H(t)\Psi^*(R,-t),
\label{eq:tr-tdse}
\ee
where the third condition, Eq.~(\ref{eq:tr-sym-pot}), is used to obtain
the right-hand-side.
Since Eq.~(\ref{eq:tdse}) and Eq.~(\ref{eq:tr-tdse}) have the same form,
the forward and backward propagations are described by
the same propagator $\hat U(t,t_0)$ as
\be
\Psi(R,t) &=& \hat U(t,t_0) \Psi(R,t_0), \label{eq:forward-wf} \\
\Psi^*(R,-t) &=& \hat U(t,t_0) \Psi^*(R,-t_0). \label{eq:backward-wf}
\ee

For simplicity, we explicitly consider the forward time-propagation from $-T/2$ to $t$ as
\be
\Psi\left(R,t-\frac{T}{2} \right) &=& \hat U\left (t-\frac{T}{2},-\frac{T}{2} \right)
\Psi \left (R, -\frac{T}{2}\right),
\label{eq:forward-wf2}
\ee
where $0\le t \le T$.
Employing the first and second conditions,
Eq.~(\ref{eq:tr-sym-wf}) and Eq.~(\ref{eq:adiabatic-wf}),
the corresponding backward time-propagation is described as
\be
\Psi^*\left(R,-t+\frac{T}{2} \right) &=& \hat U\left (t-\frac{T}{2},-\frac{T}{2} \right)
\Psi^* \left (R, \frac{T}{2}\right) \nonumber \\
&=&e^{i(\phi_1+\phi_2)}\hat U\left (t-\frac{T}{2},-\frac{T}{2} \right)
\Psi \left (R, -\frac{T}{2}\right). \nonumber \\
\label{eq:backward-wf2}
\ee

Comparing with Eq.~(\ref{eq:forward-wf2}) and Eq.~(\ref{eq:backward-wf2}),
one finds that the forward and backward-propagated wavefunctions have the following relation:
\be
\Psi^*\left(R,-t+\frac{T}{2} \right) = 
e^{i\phi}\ \Psi\left(R,t-\frac{T}{2} \right),
\label{eq:forard-backward}
\ee
where the constant phase $\phi$ is defined as $\phi\equiv \phi_1 + \phi_2$.
Therefore, the three conditions, Eq.~(\ref{eq:tr-sym-wf}), Eq.~(\ref{eq:adiabatic-wf}) 
and Eq.~(\ref{eq:tr-sym-pot}), lead to the equivalence of forward and backward
time-propagations, or namely the time-reversal symmetry of the system.

Evaluating the current flow with Eq.~(\ref{eq:tr-sym-pot})
and Eq.~(\ref{eq:forard-backward}),
the following constraint for the current is obtained:
\be
&&J\left ( t-\frac{T}{2}\right) = J^*\left ( t-\frac{T}{2}\right)  \nonumber \\
&&=\left [\int dR \Psi^*\left(R,t-\frac{T}{2} \right) 
\hat J\left ( t-\frac{T}{2} \right )
\Psi\left(R,t-\frac{T}{2} \right) \right]^*\nonumber \\
&&=-\int dR \Psi^*\left(R,-t+\frac{T}{2} \right)
\hat J\left ( -t+\frac{T}{2} \right )
\Psi\left(R,-t+\frac{T}{2} \right) \nonumber \\
&&= 
-J\left ( -t+\frac{T}{2}\right).
\label{eq:tr-sym-current}
\ee
Furthermore, the transported charge $Q$ during the time interval between $-T/2$ and $T/2$
becomes
\be
Q &=& \int^{T}_{0}dt J\left (t -\frac{T}{2} \right) 
= -\int^{T}_{0}dt J\left (-t +\frac{T}{2} \right)  \nonumber \\
&=& -\int^{T}_{0}d\tau J\left (\tau -\frac{T}{2} \right) 
=-Q
\ee
with the variable transformation, $\tau = -t +T$.
Hence the transported charge is zero, $Q=0$,
under the three conditions, Eq.~(\ref{eq:tr-sym-wf}),
Eq.~(\ref{eq:adiabatic-wf}), and Eq.~(\ref{eq:tr-sym-pot}).
Therefore, at least, one of the three conditions has to be
violated in order to induce charge-transport or direct current by
external fields.
The violation of the first condition, Eq.~(\ref{eq:tr-sym-wf}),
allows initial states to have current flow, and
it may trivially induce the charge transport.

The second condition, Eq.~(\ref{eq:adiabatic-wf}), can be violated
by photo-excitation. Therefore, the photovoltaic effect may be induced in a resonant
excitation condition, where the photon energy of applied fields exceeds
the optical gap of materials. Indeed, the shift-current mechanism relies on
the violation of the second condition as it can be induced by linearly polarized
light in a resonant condition, while satisfying the first and third conditions,
Eq.~(\ref{eq:tr-sym-wf}) and Eq.~(\ref{eq:tr-sym-pot}).

Importantly, even if applied electric fields are strongly off-resonant, 
charge-transport can be induced by a strongly-nonlinear light-matter interactions
with inversion-symmetry-breaking \cite{Schiffrin2012,PhysRevLett.113.087401,Wachter_2015}.
In Refs~\cite{Schiffrin2012,Krausz2014}, 
such current induced by linearly-polarized light has been interpreted
with reversible and adiabatic quantum transitions based on \textit{virtual} carrier generation,
and they suggested an efficient and high-speed signal processing based on the reversibility.
However, according to the above analysis, an irreversible transition violating
Eq.~(\ref{eq:adiabatic-wf}) with \textit{real} photocarrier generation is indispensable
for the current injection in dielectrics with linearly-polarized light
that satisfies Eq.~(\ref{eq:tr-sym-pot})
even if the spatial inversion symmetry is broken by crystal structures or laser pulses.
Thus the microscopic mechanism of optical-field-induced current in dielectrics
warrants further investigation.

The above analysis further indicates that
the strongly-nonlinear light-matter interactions 
break the second condition, Eq.~(\ref{eq:adiabatic-wf}),
through nonlinear photocarrier generation
such as multi-photon absorption and tunneling excitation \cite{keldysh1965ionization},
resulting in the field-induced current.
Thus, the light-induced current in dielectrics can be seen as the current
mediated by photocarriers.
Recently, the role of intraband transitions in photocarrier generation has
been discussed \cite{PhysRevB.98.035202}, and they suggested a potential
for efficient control of photocarrier generation with multicolor laser pulses
by optimizing the contribution of inter and intrabant transitions.
Extending this proposal, the photocurrent may be efficiently induced 
with multicolor laser pulses by manipulating
the contributions of inter and intraband transitions.

The third condition, Eq.~(\ref{eq:tr-sym-pot}), is the time-reversal symmetry
of the Hamiltonian, and it can be broken by a magnetic field, circularly-polarized
light etc. For example, by applying a static magnetic field, the quantum Hall effect
can be hosted in a two-dimensional electron gas
\cite{PhysRevLett.45.494,KOHMOTO1985343}.
In addition to the shift current, the injection current is yet another
mechanism of the photovoltaic effect based on the breakdown of time reversal symmetry.
The injection current originates from the photocarrier population imbalance
in momentum space induced by circularly- or elliptically-polarized light
in a system with breakdown of spatial inversion symmetry \cite{PhysRevB.61.5337}.

In summary, we theoretically investigated field-induced charge-transport phenomena
from the viewpoint of time-reversal symmetry.
We clarified that the three conditions, Eq.~(\ref{eq:tr-sym-wf}),
Eq.~(\ref{eq:adiabatic-wf}), and Eq.~(\ref{eq:tr-sym-pot}),
guarantee the identity of the forward and backward time-propagation,
Eq.~(\ref{eq:forard-backward}), and forbid the charge transport processes.
Therefore, the light-induced charge transport, or namely photovoltaic effects,
can be induced only if, at least, one of these conditions is violated
by breaking the time-reversal symmetry of the target system.
This finding provides microscopic insights into the photocurrent generation
and may open an efficient way of inducing photovoltaic effects
and controlling of electric current by light.


\begin{acknowledgments}
  The author acknowledges fruitful discussions with D.~Shin,
  H.~H\"ubener, U.~De~Giovannini, and A.~Rubio.
\end{acknowledgments}


\bibliography{ref}

\end{document}